\documentclass[12pt,preprint]{aastex}
\usepackage{natbib}

\def\Fvar{\ifmmode F_{\rm var} \else $F_{\rm var}$\fi}
\def\Rmax{\ifmmode R_{\rm max} \else $R_{\rm max}$\fi}
\def\tcent{\ifmmode \tau_{\rm cent} \else $\tau_{\rm cent}$\fi}
\def\tpeak{\ifmmode \tau_{\rm peak} \else $\tau_{\rm peak}$\fi}
\def\rmax{\ifmmode r_{\rm max} \else $r_{\rm max}$\fi}
\def\MBH{\ifmmode M_{\rm BH} \else $M_{\rm BH}$\fi}
\def\sigstar{\ifmmode \sigma_* \else $\sigma_*$\fi}
\def\sigline{\ifmmode \sigma_{\rm line} \else $\sigma_{\rm line}$\fi}
\def\Msigma{\ifmmode M_{\rm BH}$--$\sigma_* 
  \else $M_{\rm BH}$--$\sigma_*$\fi}

\def\kms{\ifmmode {\rm km\ s}^{-1} \else km s$^{-1}$\fi}
\def\Msun{\ifmmode M_{\odot} \else $M_{\odot}$\fi}
\def\Lsun{\ifmmode L_{\odot} \else $L_{\odot}$\fi}
\def\qo{\ifmmode q_{\rm o} \else $q_{\rm o}$\fi}
\def\Ho{\ifmmode H_{\rm o} \else $H_{\rm o}$\fi}
\def\ho{\ifmmode h_{\rm o} \else $h_{\rm o}$\fi}

\def\vFWHM{\ifmmode v_{\mbox{\tiny FWHM}} \else
            $v_{\mbox{\tiny FWHM}}$\fi}
\def\CCF{\ifmmode F_{\it CCF} \else $F_{\it CCF}$\fi}
\def\ACF{\ifmmode F_{\it ACF} \else $F_{\it ACF}$\fi}
\def\Halpha{\ifmmode {\rm H}\alpha \else H$\alpha$\fi}
\def\Hbeta{\ifmmode {\rm H}\beta \else H$\beta$\fi}
\def\Hgamma{\ifmmode {\rm H}\gamma \else H$\gamma$\fi}
\def\Hdelta{\ifmmode {\rm H}\delta \else H$\delta$\fi}
\def\Lya{\ifmmode {\rm Ly}\alpha \else Ly$\alpha$\fi}
\def\Lyb{\ifmmode {\rm Ly}\beta \else Ly$\beta$\fi}

\def\ciii{\ifmmode {\rm C}\,{\sc iii} \else C\,{\sc iii}\fi}
\def\civ{\ifmmode {\rm C}\,{\sc iv} \else C\,{\sc iv}\fi}

\def\o5007{[O\,{\sc iii}]\,$\lambda5007$}

\shorttitle{}
\shortauthors{Peterson et al.}

\begin{document}
\title{Erratum: ``Multiwavelength Monitoring of the Dwarf Seyfert 1
Galaxy NGC~4395. I. A Reverberation-Based 
Measurement of the Black Hole Mass'' (ApJ, 632, 799 [2005])} 
\author{
Bradley~M.~Peterson,
Misty~C.~Bentz,
Louis-Benoit~Desroches,
Alexei~V.~Filippenko,
Luis~C.~Ho,
Shai~Kaspi,
Ari~Laor,
Dan~Maoz,
Edward~C.~Moran,
Richard~W.~Pogge, and
Alice~C.~Quillen
}

In the original version of this paper, we reported on
a series of ultraviolet (UV) spectroscopic
observations of the dwarf Seyfert 1 galaxy
NGC~4395, made with the 
Space Telescope Imaging Spectrograph on
{\em Hubble Space Telescope (HST).}
Unfortunately, a data processing error led to an incorrect
flux calibration for these spectra. All STIS-based UV fluxes in
the original paper are too high by a factor of 7.96 
as a result of 
neglecting to adjust the flux-scaling algorithm from
a diffuse source to a point source.

Because most of the analysis in the original paper involved
only relative flux changes, most of the scientific conclusions
are unaffected by this correction, except
for the slope of the broad-line region radius--luminosity relationship, 
as described below.

Specific changes that result from this correction are:
\begin{enumerate}
\item All of the values for the continuum and \civ\ 
emission-line fluxes and their associated uncertainties
in the electronic Tables 1 and 2 need
to be multiplied by a factor of 0.126.
\item The mean and rms fluxes in column (3) of Table 3 need to be 
multipled by a factor of 0.126.
\item The luminosity entries for NGC~4395 in column (4) of Table 6
need to be multiplied by a factor of 0.126. A corrected version of 
Table 6 is provided here.
\item The flux scales in Figs.\ 1, 2, and 5 should be multiplied by
a factor of 0.126.
\end{enumerate}

The only substantive change resulting from this correction
is the relationship between the broad-line region radius,
as measured by the time response of the \civ\ emission line,
and the UV continuum luminosity. We show here the 
corrected version of Fig.\ 6, based on the entries in the
corrected version of Table 6. The best fit power-law
relationship to these data is
\begin{equation}
\log R_{\rm BLR}{\mbox{\rm (lt-days)}} = (1.06 \pm 0.16) + (0.61 \pm 
0.05)\log \left( \frac{L_{\rm UV}}{10^{44}\,{\rm erg\ s}^{-1}}\right),
\end{equation}
which replaces eq.\ (3) in the original paper.
The slope $\alpha = 0.61 \pm 0.05$ is in much better agreement with the 
slope of the radius--luminosity relationship of
S.\ Kaspi et al.\ (ApJ, 629, 61 [2005]) for
\Hbeta\ and the UV continuum,
$\alpha = 0.56 \pm 0.05$.

\begin{figure}
\figurenum{6}
\epsscale{1} 
\plotone{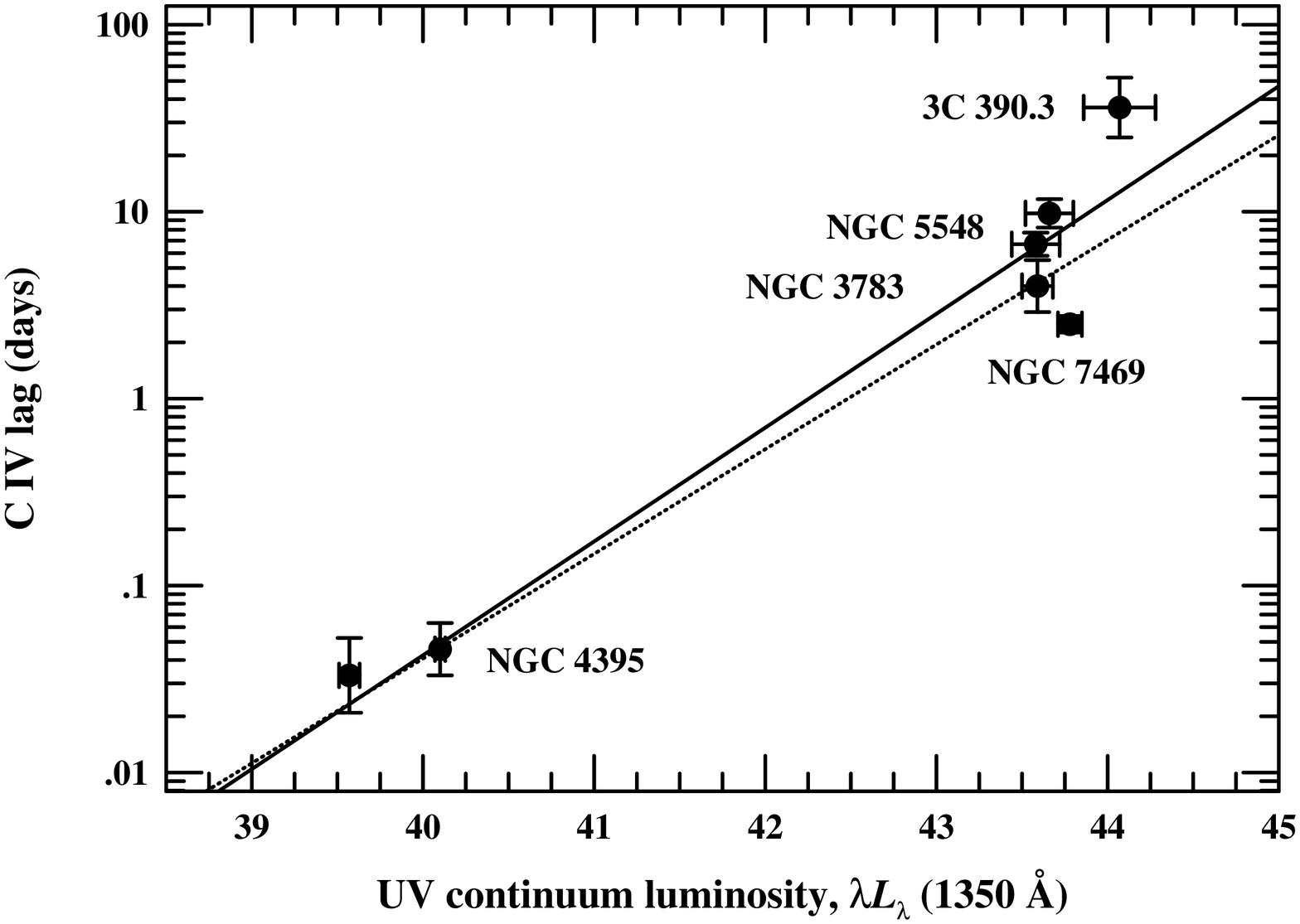} 
\caption{The radius--luminosity relationship based on
the \civ\,$\lambda1549$ emission line and the UV continuum,
for the galaxies listed in Table 6.
The UV continuum luminosity is in units of erg\,s$^{-1}$.
The best-fit line, with slope $\alpha = 0.61\pm 0.05$,
is shown as a solid line. The dashed line is the best fit
for a fixed slope $\alpha = 0.56$, which is the
slope of the relationship between the size of the
\Hbeta-emitting region and the UV luminosity
(Kaspi et al.\ 2005).}
\end{figure}  

\setcounter{table}{5}     
\begin{deluxetable}{lcccc}
%
%
\tablewidth{0pt}
\tablecaption{Measured C\,{\sc iv} $\lambda1549$ Lags}
\tablehead{
\colhead{} &
\colhead{Lag} &
\colhead{} &
\colhead{$\log \lambda L_{\lambda}({\rm UV})$} &
\colhead{} \\
\colhead{Data Set} &
\colhead{(days)} &
\colhead{Reference\tablenotemark{a}} &
\colhead{(erg\ s$^{-1}$)} &
\colhead{Reference\tablenotemark{a}} \\
\colhead{(1)} & 
\colhead{(2)} & 
\colhead{(3)} & 
\colhead{(4)} & 
\colhead{(5)} 
} 
\startdata
NGC 4395 -- Visit 2 & $0.033^{+0.017}_{-0.013}$ & 1 & $39.57 \pm 0.06$ 
& 1 \\
NGC 4395 -- Visit 3 & $0.046^{+0.017}_{-0.013}$ & 1 & $40.10 \pm 0.03$ 
& 1 \\
NGC 3783            & $4.0^{+1.0}_{-1.5}$       & 2 & $43.59 \pm 0.09$ 
& 3 \\
NGC 5548 -- Year 1  & $9.8^{+1.9}_{-1.5}$       & 2 & $43.66 \pm 0.14$ 
& 4 \\
NGC 5548 -- Year 5  & $6.7^{+0.9}_{-1.0}$       & 2 & $43.58 \pm 0.06$ 
& 5 \\
NGC 7469            & $2.5^{+0.3}_{-0.2}$       & 2 & $43.78 \pm 0.07$ 
& 6 \\
3C 390.3            & $35.7^{+11.4}_{-14.6}$    & 2 & $44.07 \pm 0.21$ 
& 7 
\enddata
\tablenotetext{a}{
1: This work (ApJ, 632, 799 [2005]);
2: Peterson et al.\ 2004 (ApJ, 613, 682);
3: Reichert et al.\ 1994 (ApJ, 425, 582);
4: Clavel et al.\ 1991 (ApJ, 366, 64);
5: Korista et al.\ 1995 (ApJS, 97, 285);
6: Wanders et al.\ 1997 (ApJS, 113, 69);
7: O'Brien et al.\ 1998 (ApJ, 509, 163).}
\end{deluxetable}

\end{document}